\documentclass[pra,twocolumn]{revtex4-1}

\usepackage{amsmath}
\usepackage{amssymb}
\usepackage{graphicx}
\usepackage{mathbbol}

\newcommand{\bra}[1]{\left\langle #1 \right|}
\newcommand{\ket}[1]{\left| #1 \right\rangle}

\begin{document}

\title{Monogamy relation in multipartite continuous-variable quantum teleportation}

\author{Jaehak Lee}
\author{Se-Wan Ji}
\author{Jiyong Park}
\author{Hyunchul Nha}
\affiliation{Department of Physics, Texas A \& M University at Qatar, P.O. Box 23874, Doha, Qatar}

\begin{abstract}
Quantum teleportation (QT) is a fundamentally remarkable communication protocol that also finds many important applications for quantum informatics. Given a quantum entangled resource, it is crucial to know to what extent one can accomplish the QT. This is usually assessed in terms of output fidelity, which can also be regarded as an operational measure of entanglement. In the case of multipartite communication when each communicator possesses a part of $N$-partite entangled state, not all pairs of communicators can achieve a high fidelity due to monogamy property of quantum entanglement. 
We here investigate how such a monogamy relation arises in multipartite continuous-variable (CV) teleportation particularly using a Gaussian entangled state. 
We show a strict monogamy relation, i.e. a sender cannot achieve a fidelity higher than optimal cloning limit with more than one receiver. While this seems rather natural owing to the no-cloning theorem, a strict monogamy relation still holds even if the sender is allowed to individually manipulate the reduced state in collaboration with each receiver to improve fidelity. The local operations are further extended to non-Gaussian operations such as photon subtraction and addition, and we demonstrate that the Gaussian cloning bound cannot be beaten by more than one pair of communicators. Furthermore we investigate a quantitative form of monogamy relation in terms of teleportation capability, for which we show that a faithful monogamy inequality does not exist.
\end{abstract}

\maketitle

\section{\label{sec:introduction}Introduction}

Quantum teleportation \cite{bib:PhysRevLett.70.1895, bib:PhysRevLett.80.869} is a fundamental communication protocol to transfer quantum information from one location to another. It can represent a distinguishing feature of quantum mechanics and also find many practical applications such as universal quantum computation \cite{Gottesman}, entanglement swapping \cite{bib:PhysRevLett.71.4287} and quantum repeaters \cite{bib:PhysRevLett.81.5932,bib:PhysRevA.59.169}. To assess the faithfulness of quantum teleportation, one usually employs output fidelity as a figure of merit, for which two important benchmarks exist, i.e., classical bound and no-cloning bound. Classical bound $F_\textrm{cl}$ is determined by the maximum fidelity achieved under classical measure-and-prepare protocols \cite{bib:PhysRevLett.74.1259, bib:PhysLettA.253.249, bib:PhysRevLett.86.4938, bib:PhysRevLett.112.010501}. If a fidelity beyond the classical bound is obtained, one can be convinced that the teleportation makes use of genuine quantum entanglement. On the other hand, the no-cloning bound $F_\textrm{nc}$ is a stricter benchmark arising from the no-cloning theorem \cite{bib:PhysRevA.58.1827, bib:PhysRevA.63.052313, bib:PhysRevA.64.010301, bib:PhysRevLett.95.070501}. If one achieves output fidelity beyond the no-cloning bound, it guarantees that there does not exist other party who possesses a copy at the same level of fidelity or higher. For CV teleportation \cite{bib:PhysRevLett.80.869, bib:Science.282.706, bib:PhysRevLett.94.220502}, where the input state is prepared as a coherent state with unknown displacement uniformly distributed in phase space, the classical bound is given by $ F_\textrm{cl} = \frac{1}{2} $ \cite{bib:PhysRevLett.86.4938}. On the other hand, the no-cloning bound is given by $ F_\textrm{nc}^G = \frac{2}{3} $ when the resources and the operations are restricted to Gaussian regime \cite{bib:PhysRevA.64.010301}, and $ F_\textrm{nc} \approx 0.6826 $ when no restriction is made to also include non-Gaussian operations \cite{bib:PhysRevLett.95.070501}. A successful teleportation beyond no-cloning limit was experimentally demonstrated \cite{bib:PhysRevLett.94.220502}.

In a variety of studies such as entanglement distillation \cite{bib:PhysRevLett.76.722, bib:PhysRevA.61.032302, bib:PhysRevA.65.062306, bib:PhysRevA.67.032314, bib:PhysRevA.84.012302, bib:PhysRevA.87.032307,Kim} and robustness of entanglement \cite{bib:PhysRevA.76.030301, bib:PhysRevA.82.052308}, teleportation fidelity has been employed as an operational measure of entanglement to test if an entangled state at hand is a useful resource. For multipartite communications, it is important to know how useful a given multipartite entangled state is in view of performance achieved individually by each pair of users and performance achieved collectively by all users. In this paper, we investigate the CV teleportation to examine the monogamy property of useful multipartite CV entanglement. Entanglement monogamy was first developed in terms of Coffman-Kundu-Wooters (CKW) inequality \cite{bib:PhysRevA.61.052306} stating that the sum of entanglement shared by several parties is restricted by the total amount of entanglement. A strict monogamy relation was later found such that if one party has quantum correlation with a certain number parties, (s)he cannot be correlated with the other parties. Such a relation is proved to be true in quantum nonlocality \cite{bib:PhysRevA.73.012112} and quantum steering \cite{bib:PhysRevA.88.062108}, particularly for Gaussian states under Gaussian measurement \cite{Ji}. However, extended to non-Gaussian measurements, such a strict monogamy may break down even for Gaussian states \cite{bib:NGsteering}. On the other hand, it was also found that the quantum dense coding protocol gives a strict monogamy relation \cite{bib:PhysRevA.87.052319, bib:PhysRevA.90.022301}, i.e., a sender cannot have quantum advantage with more than one receiver simultaneously.

It is possible for Alice to achieve quantum advantage to some extent beyond classical fidelity with any number of communicators. 
However, the no-cloning theorem naturally implies a strict monogamy in QT, as the latter belongs to a subset of all possible state-manipulations to make quantum copies considered in the no-cloning theorem. Let us assume a quantum state $ \rho $ shared by three parties, Alice, Bob and Charlie. When Alice tries to teleport an input state to Bob and Charlie simultaneously, both of the output fidelities $ F_{A:B} $ and $ F_{A:C} $ cannot beat no-cloning bound. While an optimal cloning scheme saturates the no-cloning bound, that is, $ F_{A:B} = F_{A:C} = F_\textrm{nc} $, it is not immediately obvious how one can come up with a QT scheme to accomplish the optimal cloning. We identify the CV teleportation protocol to achieve the optimal cloning both in Gaussian and non-Gaussian regime.
On the other hand, one might wonder if the no-cloning bound can be beaten when we generalize conditions on possible strategies for teleportation protocol. We may attempt to beat Gaussian cloning limit in two different ways. First one is to relax the constraint of simultaneous teleportation. Alice shares two different copies of the same quantum state $ \rho $ and each copy is used for teleportation with Bob and Charlie, respectively. They are allowed to improve teleportation fidelity by individually manipulating the reduced state with Gaussian unitary operations. We show that it is still not possible that both of the telerportation fidelities beat the Gaussian cloning bound. Another scenario is to apply non-Gaussian operations on Gaussian state. Since the genuine no-cloning bound $ F_\textrm{nc} $ is slightly higher than Gaussian one $ F_\textrm{nc}^G $, we are interested to know if $ F_\textrm{nc}^G $ can be achieved by manipulating a Gaussian entangled state with non-Gaussian operations. We demonstrate that the non-Gaussian operations such as photon subtraction and photon addition do not lead to beat the Gaussian-cloning bound. These results make a strong support to the statement that CV teleportation monogamy is a quite strict relation.

Furthermore, we make a quantitative analysis of the monogamy inequality in a form
\begin{equation} \label{eq:inequality}
E_{A:B}^\alpha+E_{A:C}^\alpha \leq E_{A:BC}^\alpha ,
\end{equation}
where $E_{A:B(C)}$ is entanglement measure between A and B (C), while $E_{A:BC}$ is between A and BC. In the case of $ \alpha = 2 $, the inequality recovers the original CKW inequality \cite{bib:PhysRevA.61.052306}, which proved that concurrence satisfies the inequality for three qubits, later generalized to $ N $ qubits \cite{bib:PhysRevLett.96.220503}. It was also shown that squashed entanglement \cite{bib:PhysRevA.69.022309} and Gaussian tangle \cite{bib:PhysRevLett.98.050503} satisfy the inequality with $ \alpha = 1 $. For discrete-variable teleportation, monogamy inequality was investigated in terms of teleportation capability $ \mathcal{C} $ \cite{bib:PhysRevA.79.054309}, which quantifies the quantum advantage in teleportation fidelity beyond classical limit. It was shown that monogamy inequality with $ \alpha = 1 $ is satisfied for any $ N $-qubit states and for three-qutrit pure states, but neither for general $ N $-qutrit states nor in higher dimensions. We investigate the monogamy inequality for CV teleportation and show that the inequality is violated for any finite $ \alpha $. We further demonstrate that teleportation capability does not obey any nontrivial monogamy inequality.

This paper is orginized as follows. In Sec. \ref{sec:optimalcloning}, we briefly review no-cloning theorem and its application to monogamy relation of quantum teleportation. We then show that CV teleportation can achieve optimal cloning. In Sec. \ref{sec:strictmonogamy}, we investigate CV teleportation with one sender and two receivers. We find that a strict monogamy still holds even if Gaussian operations or non-Gaussian operations are allowed individually for each pair of communicators to improve teleportation fidelity. In Sec. \ref{sec:monogamyinequality}, we make a quantitative analysis of monogamy inequality in terms of teleportation capability. We show that such an inequality does not hold in CV teleportation. In Sec. \ref{sec:conclusion}, we summarize our results with concluding remarks.

\section{\label{sec:optimalcloning}Teleportation and optimal cloning}

\subsection{No-cloning theorem and simultaneous teleportation}

\begin{figure}[b]
\centering \includegraphics[clip=true, width=0.8\columnwidth]{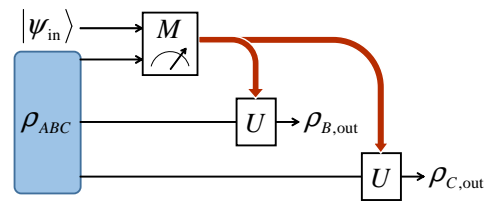}
\caption{\label{fig:simultaneous} Schematic for simultaneous teleportation. $M$ and $U$ represent measurement and unitary transformation, respectively.}
\end{figure}

Here we address the implication of the no-cloning theorem on the monogamy relation of QT. Let us consider a protocol where Alice teleports an input state to Bob and Charlie simultaneously as described in Fig. \ref{fig:simultaneous}.
In a standard teleportation protocol, Alice takes a joint measurement on her state combined with the input state and then sends her measurement outcome to Bob. Bob then obtains the output state by performing a local operation on his system according to the measurement outcome. In our simultaneous teleportation protocol, Alice sends her measurement outcome to both Bob and Charlie who performs local operations accordingly. 
It is not possible that both Bob and Charlie achieve output fidelity greater than the no-cloning bound, due to the no-cloning theorem.
It is also straightforward to generalize this argument to the case of more than two receivers, which gives a strict monogamy relation. Among $N$ receivers, only one can obtain the output state beating the no-cloning bound.

In the protocol above, sender and receivers use the distributed state without any operations before the QT protocol. We show that, even if any local trace-preserving operations are allowed to improve teleportation fidelity, a strict monogamy relation still holds for symmetric states. Let a ($ N+1 $)-partite state $\rho_{AB_1B_2 \cdots B_N}$ be distributed to Alice and $N$ Bobs, which is symmetric under permutation among different Bobs. We denote $\mathcal{S}_i$ and $\mathcal{T}_i$ trace-preserving operators applied on $A$ and $B_i$ that give an optimized fidelity $F_{A:B_i}^\textrm{opt}$ between $A$ and $B_i$, respectively. Due to symmetry, the optimization is identical for different $B_i$'s, i.e., $ \mathcal{S}_1 = \mathcal{S}_2 = \cdots = \mathcal{S}_N \equiv \mathcal{S} $, $ \mathcal{T}_1 = \mathcal{T}_2 = \cdots = \mathcal{T}_N \equiv \mathcal{T} $. Then we suppose that all parties apply their own operations simultaneously such that the state is transformed into $ \rho_{AB_1B_2 \cdots B_N}' = \left(\mathcal{S}\bigotimes \mathcal{T}\bigotimes\mathcal{T}\bigotimes\cdots\bigotimes\mathcal{T}\right) \rho_{AB_1B_2 \cdots B_N} $. Now $\rho_{AB_1B_2 \cdots B_N}'$ is another quantum state which yields the optimized teleportation fidelity $F_{A:B_i}^\textrm{opt}$ for each $B_i$ and all $F_{A:B_i}^\textrm{opt}$'s are the same due to symmetry. Therefore, owing to the no-cloning theorem, we must have $ F_{A:B_1}^\textrm{opt} = F_{A:B_2}^\textrm{opt} = \cdots = F_{A:B_N}^\textrm{opt} \leq F_\textrm{nc} $.

\subsection{Optimal $ 1 \to N $ cloning}

Although it is rather obvious that a simultaneous teleportation cannot achieve the fidelity beyond no-cloning limit, it is not straightforward to see whether there exists a QT scheme to achieve the optimal cloning, for QT is only a subset of all possible state manipulations. We here come up with a CV teleportation protocol achieving the optimal cloning fidelity with an appropriate entangled state, both in Gaussian and non-Gaussian regime. 
The optimal $ 1 \to N $ cloning of coherent states was initially investigated within Gaussian regime \cite{bib:PhysRevA.64.010301} to give the so-called Gaussian cloning bound. It was later shown in \cite{bib:PhysRevLett.95.070501} that a non-Gaussian resource can slightly improve the fidelity to give the ultimate cloning bound. Ref. \cite{bib:PhysRevLett.95.070501} also explicitly showed a method to achieve the optimal cloning, which we briefly review. We then propose a teleportation protocol that leads to the same output states.

Given a $ 1 \to N $ cloning transformation $\boldsymbol{\mathcal{T}}$, the fidelity between input state and $j$th output state can be written by
\begin{equation}
F_j = \textrm{Tr} \left[ \boldsymbol{\mathcal{T}} \left( \rho_\textrm{in} \right) \left( \mathbb{1}\otimes\cdots\otimes\mathbb{1} \otimes\rho_\textrm{in}\otimes \mathbb{1}\otimes\cdots\otimes\mathbb{1} \right) \right],
\end{equation}
where  $\mathbb{1}$ is identity operator. 
With the covariant property of the optimal cloning, the transformation $\boldsymbol{\mathcal{T}}$ is described by
\begin{equation}
\chi_\textrm{out} \left( \xi_1, \xi_2, \cdots, \xi_N \right) = t \left( \xi_1, \xi_2, \cdots, \xi_N \right) \chi_\textrm{in} \left( \textstyle \sum_i \xi_i \right) ,
\end{equation}
where $ \chi_\textrm{out} \left( \xi_1, \xi_2, \cdots, \xi_N \right) = \textrm{Tr} \left[ \rho_\textrm{out} W_{\xi_1, \xi_2, \cdots, \xi_N} \right] $ and $ \chi_\textrm{in} \left( \xi \right) = \textrm{Tr} \left[ \rho_\textrm{in} W_\xi \right] $ are characteristic functions of output and input states respectively, with $ W_\xi $ the Weyl operator in a relevant Hilbert space. The multiplicative function $ t \left( \xi_1, \xi_2, \cdots, \xi_N \right) $ is given by the characteristic function of a state $ \rho_T $ under a suitable linear transformation $ \Omega $, i.e., $
t \left( \xi_1, \xi_2, \cdots, \xi_N \right) = \textrm{Tr} \left[ \rho_T W_{\Omega \left( \xi_1, \xi_2, \cdots, \xi_N \right)} \right] $ .
We here consider symmetric cloning so that we maximize the fidelity $ \frac{1}{N} \sum_j F_j $ . The problem then reduces to optimizing over a certain quantum state $ \rho_T $ to maximize the fidelity
\begin{equation} \label{eq:clonefidelity}
\frac{1}{N} \sum_{j=1}^N F_j = \frac{1}{N} \textrm{Tr} \left[ \rho_T \sum_{j=1}^N \exp \left( -\frac{\hat{P}_j^2 + ( \sum_{k\neq j} \hat{Q}_k )^2}{2} \right) \right],
\end{equation}
where $\{\hat{P}_j,\hat{Q}_j\}(j=1,2,\cdots,N)$ is a set of legitimate field operators in the $N$-mode Hilbert space.
Although an analytical optimization is not straightforward, the solution can be numerically obtained. For example, $ F_\textrm{nc} \approx 0.6826 $ was obtained for $ 1 \to 2 $ cloning, which demonstrates the optimality of non-Gaussian cloning beyond Gaussian cloning limit $ F_\textrm{nc}^\textrm{G} = \frac{2}{3} $.

\begin{figure}[t]
\centering \includegraphics[clip=true, width=0.9\columnwidth]{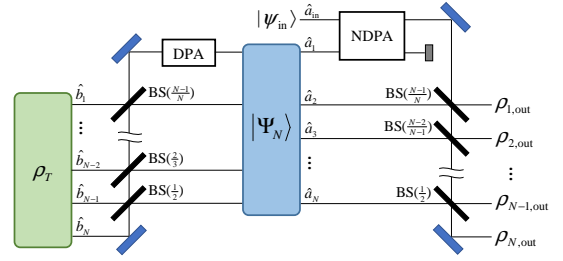}
\caption{\label{fig:cloner} A scheme for $ 1 \to N $ cloning. DPA and NDPA represent degenerate and non-degenerate parametric amplifier, respectively. $ \textrm{BS}(T) $ is a beam splitter with transmittance $T$.}
\end{figure}

Now we show that the scheme of Fig. \ref{fig:cloner} achieves the optimal cloning with the fidelity in Eq. (\ref{eq:clonefidelity}).
We first start with the state $ \rho_T $ and the corresponding field operators $\hat{b}_j \equiv \frac{1}{\sqrt{2}}(\hat{Q}_j+i\hat{P}_j) $ $(j=1,2,\cdots,N)$. The transformation from $ \rho_T $ to $ \ket{\Psi_N} $ is described in the Heisenberg picture as
\begin{eqnarray}
\hat{a}_1 & = & \frac{1}{\sqrt{N}} \sum_{k=1}^N \left( \frac{N}{2\sqrt{N-1}}\hat{b}_k + \frac{N-2}{2\sqrt{N-1}}\hat{b}_k^\dagger \right) \nonumber \\
\hat{a}_j & = & \sqrt{\frac{N-j+1}{N-j+2}}\hat{b}_{j-1} \nonumber \\
& & - \sqrt{\frac{1}{(N-j+1)(N-j+2)}} \sum_{k=j}^N \hat{b}_k \nonumber \\
& & \qquad \textrm{for } j = 2,3,\dots,N .
\end{eqnarray}
The input state $ \ket{\psi_\textrm{in}} $ is then amplified via non-degenerate parametric amplifier (NDPA) mixed with the first mode of $ \ket{\Psi_N} $. One output of NDPA is discarded and the other output is mixed with the rest modes of $ \ket{\Psi_N} $. Finally, the output states are described by the Heisenberg picture operators
\begin{eqnarray} \label{eq:cloneroutput}
\hat{a}_{j,\textrm{out}} & = & \hat{a}_\textrm{in} + \sqrt{\frac{N-1}{N}}\hat{a}_1^\dagger - \sqrt{\frac{N-j}{N-j+1}}\hat{a}_{j+1} \nonumber \\
& & + \sum_{k=2}^j\sqrt{\frac{1}{(N-k+2)(N-k+1)}}\hat{a}_k \nonumber \\
& = & \hat{a}_\textrm{in} - \frac{1}{\sqrt{2}}\hat{P}_j + \frac{i}{\sqrt{2}}\sum_{k\neq j}\hat{Q}_k .
\end{eqnarray}
To evaluate the output fidelity, it suffices to consider a vacuum state as input under a covariant scheme, which gives the same fidelity regardless of the amplitude of the input state. Thus the fidelity of $j$'th output is given by
\begin{eqnarray} \label{eq:outputfidelity}
F_j & = & \textrm{Tr} \left[ \ket{0}\bra{0}_j \rho_\textrm{out} \right] \nonumber \\
& = & \textrm{Tr} \left[ :\exp\left(-\hat{a}_j^\dagger\hat{a}_j\right): \rho_\textrm{out} \right]
= \left\langle :\exp\left(-\hat{a}_{j,\textrm{out}}^\dagger\hat{a}_{j,\textrm{out}}\right): \right\rangle \nonumber \\
& = & \textrm{Tr} \left[ : \exp\left(-\hat{a}_\textrm{in}^\dagger\hat{a}_\textrm{in}\right) \exp\left(-\frac{\hat{P}_j^2+(\sum_{k\neq j}\hat{Q}_k)^2}{2}\right) : \rho_T \right] \nonumber \\
& = & \textrm{Tr} \left[ \exp\left(-\frac{\hat{P}_j^2+(\sum_{k\neq j}\hat{Q}_k)^2}{2}\right) \rho_T \right] ,
\end{eqnarray}
where $:\hat{A}:$ represents the normal ordering of $\hat{A}$.
We now see that the fidelities in Eqs. (\ref{eq:outputfidelity}) and (\ref{eq:clonefidelity}) are identical, so achieving an optimal cloning fidelity corresponds to an appropriate choice of the state $ \rho_T $, with or without restriction to Gaussian states.

\subsection{Optimal cloning by CV teleportation}

\begin{figure}[t]
\centering \includegraphics[clip=true, width=0.8\columnwidth]{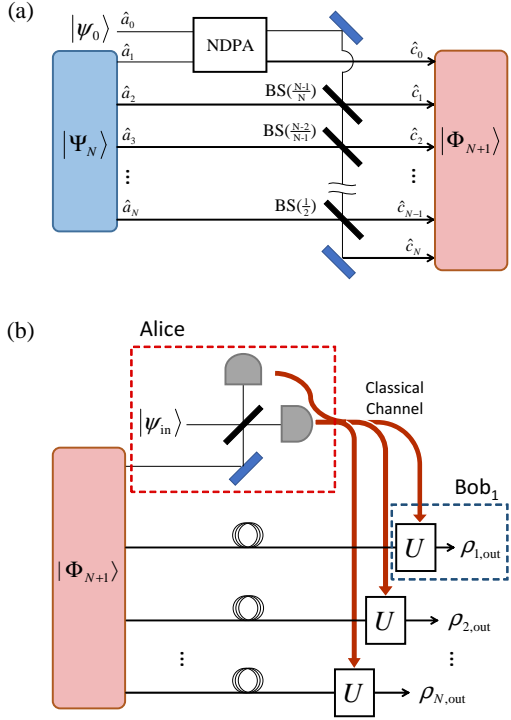}
\caption{\label{fig:optteleportation} (a) Preparation of $(N+1)$-mode state for teleportation. NDPA represents a non-degenerate parametric amplifier with gain $\frac{N}{N-1}$. $ \textrm{BS}(T) $ represents a beam splitter with transmittance $T$. (b) $ 1 \to N $ teleportation scheme.}
\end{figure}

We here demonstrate that there exists a QT scheme to produce $N$ quantum copies with optimal cloning fidelity. 
To achieve $ 1 \to N $ teleportation, we need an $(N+1)$-mode state where the first mode belongs to a sender and the other modes to $N$ receivers. We prepare the resource state $\ket{\Phi_{N+1}}$ as shown in Fig. \ref{fig:optteleportation}(a).
The Heisenberg field operators $\hat{c}_j$ $(j=0,1,\cdots,N)$ of the resource state $\ket{\Phi_{N+1}}$ can be related to $\hat{a}_j$ ($j=1,2,\cdots,N$) of the state $\ket{\Psi_N}$ and an ancilla mode $\hat{a}_0$ of state $\ket{\psi_0}$ for $j=0$ as
\begin{eqnarray}
\hat{c}_0 & = & \sqrt{\frac{1}{N-1}}\hat{a}_0^\dagger - \sqrt{\frac{N}{N-1}}\hat{a}_1 \nonumber \\
\hat{c}_j & = & \sqrt{\frac{1}{N-1}}\hat{a}_0 - \sqrt{\frac{1}{N(N-1)}}\hat{a}_1^\dagger - \sqrt{\frac{N-j}{N-j+1}}\hat{a}_{j+1} \nonumber \\
& & + \sum_{k=2}^j\sqrt{\frac{1}{(N-k+2)(N-k+1)}}\hat{a}_k \nonumber \\
& & \qquad \textrm{for } j = 1,2,\dots,N .
\end{eqnarray}
Now we show that the teleportation scheme employing the resource state $\ket{\Phi_{N+1}}$ in Fig. \ref{fig:optteleportation}(b) gives the same output states as the cloning scheme with the resource state $\ket{\Psi_N}$ in Fig. \ref{fig:cloner}. In the teleportation protocol, Alice possesses a mode $\hat{c}_0$ and $N$ Bobs the other $N$ modes. Alice combines an unknown input state with her mode and measures two quadratures $\frac{1}{\sqrt{2}}(\hat{x}_\textrm{in}-\hat{x}_{c_0})$ and $\frac{1}{\sqrt{2}}(\hat{p}_\textrm{in}+\hat{p}_{c_0})$. The measurement outcomes are sent to receivers simultaneously. Each Bob displaces his mode according to the measurement outcomes, then the output state is described by
\begin{eqnarray} \label{eq:teleportationoutput}
\hat{c}_{j,\textrm{out}} & = & \hat{c}_j + \frac{1}{\sqrt{2}}(\hat{x}_\textrm{in}-\hat{x}_{c_0}) + \frac{i}{\sqrt{2}}(\hat{p}_\textrm{in}+\hat{p}_{c_0}) \nonumber \\
& = & \hat{c}_j + \hat{a}_\textrm{in} - \hat{c}_0^\dagger \nonumber \\
& = & \hat{a}_\textrm{in} + \sqrt{\frac{N-1}{N}}\hat{a}_1^\dagger - \sqrt{\frac{N-j}{N-j+1}}\hat{a}_{j+1} \nonumber \\
& & + \sum_{k=2}^j\sqrt{\frac{1}{(N-k+2)(N-k+1)}}\hat{a}_k .
\end{eqnarray}
We see that the output states produced by teleportation are equivalent to those obtained by cloner as described in Eq. (\ref{eq:cloneroutput}).

As a remark, we note that the state $ \ket{\psi_0} $ of ancilla mode $\hat{a}_0$ in Eq. (\ref{eq:teleportationoutput}) does not affect the output state. Even though it is a highly mixed state, we obtain the same output. A crucial element when constructing the resource state $\ket{\Phi_{N+1}}$ is the gain $ g = \frac{N}{N-1} $ of NDPA in Fig. \ref{fig:optteleportation}(a), which is fully determined by $N$. It explains why we need an infinitely squeezed state for $ 1 \to 1 $ perfect teleportation. On the other hand, for $ N \geq 2 $ outputs, we achieve the optimal $ 1 \to N $ teleportation with a finitely entangled state. In the case of $N=2$, the optimal Gaussian cloning is achieved with $ \ket{\Psi_2^\textrm{G}} = \ket{0}\otimes\ket{0} $. The choice of $\ket{\psi_0}$ is not restricted and a different pure Gaussian state results in a different $\ket{\Phi_3^G}$, but all these are equivalent under local unitary operations. On the other hand, the ultimate no-cloning bound $ F_\textrm{nc} \approx 0.6826 $ is achieved by a non-Gaussian state $\ket{\Psi_2}$, which can be obtained numerically by optimizing Eq. (\ref{eq:clonefidelity}). Accordingly the optimal resource $ \ket{\Phi_3} $ for teleportation is also non-Gaussian.

\section{\label{sec:strictmonogamy}Strict monogamy relation in CV teleportation}

In the previous section, we have shown that a strict monogamy relation naturally emerges due to the no-cloning theorem. On the other hand, it is possible to improve individual teleportation fidelity for each pair if a sender and a receiver apply local operations before carrying out teleportation. For example, in the Gaussian regime, local Gaussian completely positive maps that lead to optimized fidelity was investigated \cite{bib:PhysRevA.78.062340}. It was also shown that non-Gaussian operations such as photon subtraction on two-mode squeezed vacuum state (TMSV) can improve teleportation fidelity beyond no-cloning limit even though the initial state cannot beat the no-cloning limit \cite{bib:PhysRevA.65.062306}. We here employ a three-mode Gaussian state as an initial resource state and investigate whether we can beat Gaussian no-cloning limit $F_\textrm{nc}^\textrm{G}$ by applying some local operations. We show that a strict monogamy relation still holds in the following cases.

\subsection{Improving fidelity via local Gaussian unitaries}

\subsubsection{Teleportation fidelity}

We first consider local Gaussian unitaries to improve output fidelity. A $N$-mode Gaussian state is fully described by its first and second moments of their quadrature operators $ \hat{\boldsymbol{R}} = \left( \hat{x}_1, \hat{p}_1, \hat{x}_2, \hat{p}_2, \cdots, \hat{x}_N, \hat{p}_N \right) $. The first-order moments represent the average amplitudes of field operators that can be adjusted by local displacement operations. A covariance matrix (CM) representing the second-order moments has the matrix elements given by $ \sigma_{ij} \equiv \frac{1}{2} \langle \Delta\hat{R}_i \Delta\hat{R}_j + \Delta\hat{R}_j \Delta\hat{R}_i \rangle $ where $ \Delta\hat{R}_i = \hat{R}_i - \langle \hat{R}_i \rangle $. From now on, we assume the first moments to be zero, $ \langle \hat{R}_i \rangle = 0 $, which does not affect the fidelity of covariant teleportation scheme. Let us consider a coherent state teleportation when Alice and Bob share a two-mode Gaussian state with CM in a block form
\begin{equation}
\boldsymbol{\sigma} = \left( \begin{array}{cc}
\boldsymbol{A} & \boldsymbol{C} \\
\boldsymbol{C}^T & \boldsymbol{B}
\end{array} \right) ,
\end{equation}
where $\boldsymbol{A}$, $\boldsymbol{B}$, and $\boldsymbol{C}$ are $2\times2$ real matrices. In this case, the teleportation fidelity is given by \cite{bib:RevModPhys.84.621,bib:PhysRevA.66.012304,Nha}
\begin{eqnarray}
& & F = \frac{1}{\sqrt{\det\boldsymbol{\Gamma}}} , \\
& & \boldsymbol{\Gamma} \equiv 2\boldsymbol{\sigma}_\textrm{in} + \boldsymbol{Z}\boldsymbol{A}\boldsymbol{Z} + \boldsymbol{B} - \boldsymbol{Z}\boldsymbol{C} - \boldsymbol{C}^T\boldsymbol{Z}^T , \nonumber
\end{eqnarray}
where $ \boldsymbol{\sigma}_\textrm{in} \equiv \frac{1}{2}\textrm{diag}(1,1) $ is the CM of input coherent state and $ \boldsymbol{Z} \equiv \textrm{diag}(1,-1) $. We can rewrite the fidelity in terms of the second moments of correlated quadratures $ x_- \equiv \frac{1}{\sqrt{2}}(x_1-x_2) $ and $ p_+ \equiv \frac{1}{\sqrt{2}}(p_1+p_2) $ as
\begin{equation} \label{eq:fidelity}
F = \frac{1}{\sqrt{1 + 2\left(\left\langle x_-^2 \right\rangle + \left\langle p_+^2 \right\rangle\right) + 4\left(\left\langle x_-^2 \right\rangle \left\langle p_+^2 \right\rangle - \left\langle x_- p_+ \right\rangle^2\right)}} .
\end{equation}
Here the quantities $\left\langle x_-^2 \right\rangle$, $\left\langle p_+^2 \right\rangle$, and $\left\langle x_- p_+ \right\rangle$ are not invariant under local unitary operations so the fidelity can be modified via those operations. 

Let us first assume that the cross term $ \left\langle x_- p_+ \right\rangle $ is zero and later this assumption will be justified in several cases. We then have
\begin{eqnarray}
F & = & \frac{1}{\sqrt{1 + 2\left(\left\langle x_-^2 \right\rangle + \left\langle p_+^2 \right\rangle\right) + 4\left\langle x_-^2 \right\rangle \left\langle p_+^2 \right\rangle}} \nonumber \\
& \leq & \frac{1}{\sqrt{1 + 4\sqrt{\left\langle x_-^2 \right\rangle \left\langle p_+^2 \right\rangle} + 4\left\langle x_-^2 \right\rangle \left\langle p_+^2 \right\rangle}} \nonumber \\
& = & \frac{1}{1 + 2\sqrt{\left\langle x_-^2 \right\rangle \left\langle p_+^2 \right\rangle}}.
\end{eqnarray}
Given a certain fidelity bound $f$ to overcome, the necessary condition for $ F > f $ turns out to be
\begin{equation} \label{eq:fcondition}
\left\langle x_-^2 \right\rangle \left\langle p_+^2 \right\rangle < \left(\frac{1-f}{2f}\right)^2 .
\end{equation}
For the case of the classical bound $f\equiv F_\textrm{cl} = \frac{1}{2}$, the inequality becomes the entanglement detection criterion \cite{bib:PhysRevA.60.2752}. Since one party is able to share entanglement with many different parties while the amount of each bipartite entanglement is restricted, there is no restriction on the number of parties which can obtain teleportation fidelity beyond the classical bound. For example, for an ($ N+1 $)-mode Gaussian state, which was investigated for teleportation network \cite{bib:PhysRevLett.84.3482}, Alice is entangled with $ N $ different Bobs, respectively, and achieves teleportation fidelity beyond classical bound individually with each Bob (see Section \ref{sec:monogamyinequality}).

For the Gaussian no-cloning bound $f\equiv F_\textrm{nc}^{G} = \frac{2}{3} $, we have a stricter condition
\begin{equation} \label{eq:criterion}
\left\langle x_-^2 \right\rangle \left\langle p_+^2 \right\rangle < \frac{1}{16} .
\end{equation}
A similar inequality was obtained for the quantum dense coding in \cite{bib:PhysRevA.90.022301}, which corresponds to beating the single-mode squeezed-state communication scheme.

\subsubsection{Three-mode pure state}

As a quantum resource for CVQT, let us consider a three-mode pure Gaussian state described by a CM in the standard form as \cite{bib:PhysRevA.73.032345}
\begin{equation} \label{eq:cmstandard3}
\boldsymbol{\sigma}_{ABC}^\textrm{(s)} =
\left( \begin{array}{cccccc}
a_1 & 0 & e_{12}^+ & 0 & e_{13}^+ & 0 \\ 0 & a_1 & 0 & e_{12}^- & 0 & e_{13}^- \\
e_{12}^+ & 0 & a_2 & 0 & e_{23}^+ & 0 \\ 0 & e_{12}^- & 0 & a_2 & 0 & e_{23}^- \\
e_{13}^+ & 0 & e_{23}^+ & 0 & a_3 & 0 \\ 0 & e_{13}^- & 0 & e_{23}^- & 0 & a_3
\end{array} \right) ,
\end{equation}
where off-diagonal elements $e_{ij}^{\pm}$ are fully determined by the parameters $a_i$. The three parameters $a_i$ must satisfy a triangular inequality due to the uncertainty principle as
\begin{equation} \label{eq:triangularcoef}
\left| c_2-c_3 \right| \leq 1 \leq c_2+c_3 , \textrm{ where } c_j = \frac{a_j-\frac{1}{2}}{a_1-\frac{1}{2}} \textrm{ for } j=2,3.
\end{equation}
In the standard form, there are two squeezed quadratures $ x_-^{\scriptscriptstyle AB} \equiv \frac{1}{\sqrt{2}} \left( x_1 - x_2 \right) $ and $ p_+^{\scriptscriptstyle AB} \equiv \frac{1}{\sqrt{2}} \left( p_1 + p_2 \right) $ shared between Alice and Bob, with the cross term $ \left\langle x_-^{\scriptscriptstyle AB} p_+^{\scriptscriptstyle AB} \right\rangle=0 $. Similarly we find $ x_-^{\scriptscriptstyle AC} $ and $ p_+^{\scriptscriptstyle AC} $ shared between Alice and Charlie. While operators  $\hat{x}_-$ and $\hat{p}_+$ commute for each pair, we find non-commuting operators $ \left[ \hat{x}_-^{\scriptscriptstyle AB} , \hat{p}_+^{\scriptscriptstyle AC} \right] = \frac{1}{2} \left[ \hat{x}_1 , \hat{p}_1 \right] = \frac{i}{2} $, and similarly $ \left[ \hat{x}_-^{\scriptscriptstyle AC} , \hat{p}_+^{\scriptscriptstyle AB} \right] = \frac{i}{2} $. By means of Heisenberg's uncertainty principle, the correlated quadratures must then satisfy an inequality
\begin{eqnarray} \label{eq:quadraturemonogamy}
& & \left\langle \left( \hat{x}_-^{\scriptscriptstyle AB} \right)^2 \right\rangle \left\langle \left( \hat{p}_+^{\scriptscriptstyle AB} \right)^2 \right\rangle \left\langle \left( \hat{x}_-^{\scriptscriptstyle AC} \right)^2 \right\rangle \left\langle \left( \hat{p}_+^{\scriptscriptstyle AC} \right)^2 \right\rangle \nonumber \\
& & \geq \left(\frac{1}{4}\right)^2 \left| \left\langle \left[ \hat{x}_-^{\scriptscriptstyle AB}, \hat{p}_+^{\scriptscriptstyle AC} \right] \right\rangle \right|^2 \times \left| \left\langle \left[ \hat{p}_+^{\scriptscriptstyle AB}, \hat{x}_-^{\scriptscriptstyle AC} \right] \right\rangle \right|^2 \nonumber \\
& & = \left(\frac{1}{16}\right)^2 .
\end{eqnarray}
The inequality implies that the criterion (\ref{eq:criterion}) can not be satisfied for both pairs $\{A,B\}$ and $\{A,C\}$ simultaneously, which is another proof of the strict monogamy relation.

On the other hand, the standard form (\ref{eq:cmstandard3}) might not be an optimal form which individually maximizes teleportation fidelity between Alice and Bob (Charlie). Fidelity can be improved via local Gaussian unitary operation on each mode. For now, we do not consider simultaneous teleportation. Alice shares two-mode states $ \rho_{AB} $ with Bob and $ \rho_{AC} $ with Charlie, which are reduced states of $ \rho_{ABC} $ respectively. She carries out teleportation with Bob and Charlie independently by taking different optimization operations for two different reduced states. As a consequence, the two optimized states $ \tilde{\rho}_{AB} $ and $ \tilde{\rho}_{AC} $ may not be the reduced states of a single three mode state. In this sense, this individual optimization is more generalized than the usual multipartite setting of monogamy study.

\begin{figure}[t]
\centering \includegraphics[clip=true, width=0.7\columnwidth]{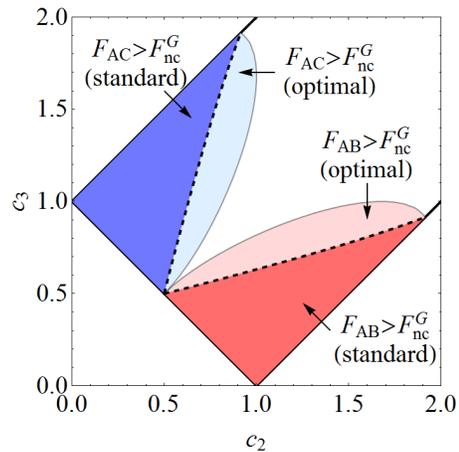}
\caption{\label{fig:monogamy} Plot illustrating the region where teleportation beats the no-cloning bound. The region colored in red (blue) is for $ F_{AB} (F_{AC}) > F_\textrm{nc}^\textrm{G} $, where $ F_{AB} (F_{AC}) $ denotes teleportation fidelity between Alice and Bob (Charlie). The dashed curves represent the boundary within which $ F_{AB} (F_{AC}) = F_\textrm{nc}^\textrm{G} $ is achieved without any local operations. Thick lines show the boundary for physical states [Ineq. (\ref{eq:triangularcoef})]. We fix $ a_1 = \frac{3}{2} $, and $a_2$ and $a_3$ are determined by $c_2$ and $c_3$ where $ c_i = \frac{a_i-1/2}{a_1-1/2} $. }
\end{figure}

When we take local Gaussian unitaries to improve fidelity, it is reasonable to consider local squeezing operation along $x$ or $p$ direction only. This is because CV teleportation makes use of two squeezed quadratures $ x_- $ and $ p_+ $, and particularly in the standard form, we do not need any phase rotation or local squeezing along other axes. The squeezing parameter can be chosen independently for each party. Specifically, Alice may choose different degree of squeezing when she carries out teleportation with Bob and Charlie, respectively.  In \cite{bib:JModOpt.50.801}, it was investigated how teleportation fidelity can be improved via local squeezing.
We show, in Fig. \ref{fig:monogamy}, the region where teleportation beats Gaussian the no-cloning bound $F_\textrm{nc}^\textrm{G}$, for which
we have numerically found the optimized fidelity by changing squeezing parameters. We see that the parameter region where the teleportation beats no-cloning bound becomes broaded with optimization. Nonetheless, we reconfirm the monogamy relation as there is no overlap between two different regions corresponding to different receivers. A tight bound is achieved, that is, $ F_{AB} = F_{AC} = F_\textrm{nc}^\textrm{G} $, only when $ a_1 = \frac{3}{2} $ and $ c_2 = c_3 = \frac{1}{2} $. The state corresponding to these parameters is exactly the same as $ \ket{\Phi_3^G} $ that achieves the optimal cloning as discussed in the previous section.

To show the strict monogamy rigorously, we recall the inequality (\ref{eq:quadraturemonogamy}). This relation still holds even though local squeezing operations are made on each pair of $ \rho_{AB} $ and $ \rho_{AC} $ independently (See also Appendix of \cite{bib:PhysRevA.90.022301} for a similar proof). It means that a strict monogamy relation is satisfied even though the simultaneous teleportation is abandoned and optimization is made independently.

\subsubsection{Mixed state}

When the resource state is pure, we can always eliminate every $ x - p $ cross-correlation term in its standard form (\ref{eq:cmstandard3}). On the other hand, it cannot be done for a general mixed state \cite{bib:PhysRevA.73.032345}. In this case, the assumption $ \left\langle x_- p_+ \right\rangle = 0 $ is no longer valid. However, in several important cases, we can remove the cross term with local Gaussian unitaries. For example, when a three-mode pure Gaussian state prepared in the standard form is distributed to each party under phase-insensitive Gaussian channel, $ x - p $ correlation term is zero. That is because $ x - p $ correlation term is already eliminated by preparing it in the standard form and phase-insensitive channel does not create such a correlation.

Even a nonzero $ \left\langle x_- p_+ \right\rangle $ can be removed by proper phase rotation for an arbitrary two-mode state. Consider a phase rotation described by
\begin{eqnarray} \label{eq:rotation}
\hat{x}_{i}' & = & \hat{x}_{i}\cos\theta_i + \hat{p}_{i}\sin\theta_i , \nonumber \\
\hat{p}_{i}' & = & \hat{p}_{i}\cos\theta_i - \hat{x}_{i}\sin\theta_i , \quad \textrm{for } i=1,2 ,
\end{eqnarray}
with $ \theta_1 = \theta $ and $ \theta_2 = -\theta $. It transforms the second moments as
\begin{eqnarray} \label{eq:rotationvariance}
\left\langle (x_-')^2 \right\rangle & = & \langle x_-^2 \rangle\cos^2\theta + \langle p_+^2 \rangle\sin^2\theta+\sin2\theta \langle x_- p_+ \rangle, \nonumber \\
\left\langle (p_+')^2 \right\rangle & = & \langle p_+^2 \rangle\cos^2\theta + \langle x_-^2 \rangle\sin^2\theta-\sin2\theta \langle x_- p_+ \rangle , \nonumber \\
\left\langle x_-' p_+' \right\rangle & = & \langle x_- p_+ \rangle\cos2\theta+\frac{1}{2} \left( \langle p_+^2 \rangle - \langle x_-^2 \rangle \right) \sin 2\theta .
\end{eqnarray}
It is readily seen that the two quantities $\langle x_-^2 \rangle + \langle p_+^2 \rangle$ and $\langle x_-^2 \rangle \langle p_+^2 \rangle - \langle x_- p_+ \rangle^2$ are invariant under the transformation (\ref{eq:rotationvariance}), and thus fidelity (\ref{eq:fidelity}) is also invariant. We can always find an angle $\theta_c$ which leads to $ \left\langle \Delta x_-' \Delta p_+' \right\rangle = 0 $, where the angle is given by
\begin{equation}
\tan2\theta_c = \frac{2\langle x_- p_+ \rangle} {\langle p_+^2 \rangle - \langle x_-^2 \rangle}.
\end{equation}
In the case when the angle $\theta_c$ is the same for both pairs $ \{ A,B \} $ and $ \{ A,C \} $, the cross term can be eliminated simultaneously. We also showed that, in this case, the inequality (\ref{eq:quadraturemonogamy}) still holds (see Appendix of \cite{bib:PhysRevA.90.022301} for proof), that is, a strict monogamy relation is satisfied.

\subsection{Non-Gaussian manipulation}

Many studies have shown that teleportation fidelity can be improved by non-Gaussian manipulation on Gaussian entangled states \cite{bib:PhysRevA.61.032302, bib:PhysRevA.65.062306, bib:PhysRevA.67.032314, bib:PhysRevA.84.012302, bib:PhysRevA.87.032307, Kim}. In particular, it was shown that a two-mode Gaussian state which does not beat no-cloning limit can overcome it by applying non-Gaussian operations \cite{bib:PhysRevA.65.062306}. Here we study whether non-Gaussian operations such as photon subtraction and addition on a three-mode Gaussian state can be used to beat Gaussian cloning limit $ F_\textrm{nc}^\textrm{G} $, while it is obvious that the ultimate no-cloning bound $ F_\textrm{nc} $ cannot be beaten.

\begin{figure}[t]
\centering \includegraphics[clip=true, width=0.7\columnwidth]{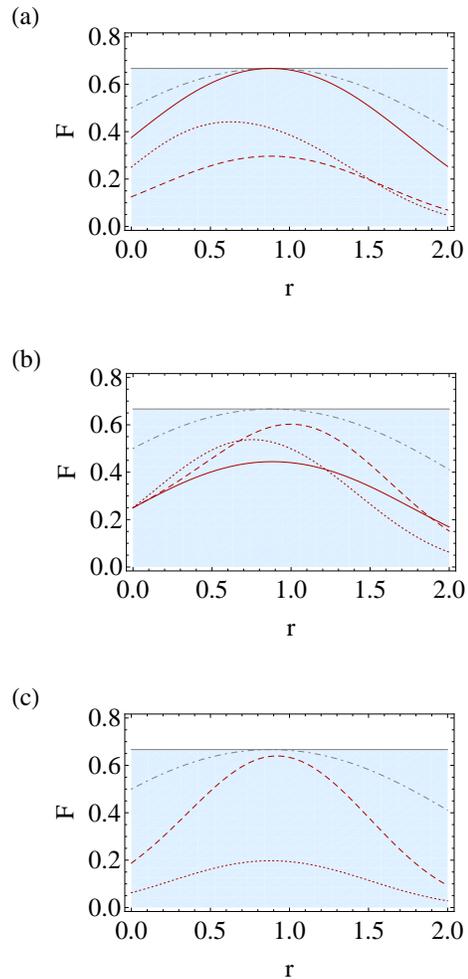}
\caption{\label{fig:nonGaussian} Plot illustrating teleportation fidelity after non-Gaussian operations. The shaded region represents the fidelity below Gaussian cloning limit $\frac{2}{3}$. Gray dot-dashed curve represents fidelity achieved with initial Gaussian state. (a) Photon subtraction operation on $A$ (solid), on $B,C$ (dashed), and on $A,B,C$ (dotted).  (b) Photon addition operation on $A$ (solid), on $B,C$ (dashed), and on $A,B,C$ (dotted). (c) Subtraction on $A$, addition on $B,C$ (dashed), and addition on $A$, subtraction on $B,C$ (dotted). }
\end{figure}

A TMSV with squeezing parameter $ r $ is prepared and its first mode is then distributed to Alice and the other modes to Bob and Charlie after dividing the second mode at a 50:50 beam splitter. The choice of $ r = \frac{1}{2}\cosh^{-1}3 $ corresponds to $ \ket{\Phi_3^G} $ achieving the optimal Gaussian cloning, but here we leave it as a free parameter. Now each party chooses one of their manipulation from photon subtraction, addition, or nothing. Since we investigate whether no-cloning limit can be beaten for both Bob and Charlie simultaneously, we assume that Bob and Charlie choose the same operation thus the same fidelity as well. In Fig. \ref{fig:nonGaussian}, we plot the fidelity after non-Gaussian operations.
Unfortunately, none of the considered cases beat Gaussian cloning limit. Moreover, fidelity does not improve compared to the original Gaussian one. The only case which leads to at least the same fidelity is observed when subtraction is made on mode $A$ with squeezing parameter $ r = \frac{1}{2}\cosh^{-1}3 $. Although photon subtraction and addition are subset of non-Gaussian operations, they are fundamental resources in quantum information and feasible in laboratory. This result indicates that a strict monogamy with Gaussian resources is not readily violated with simple non-Gaussian manipulations.

\section{\label{sec:monogamyinequality}Monogamy inequality for CV teleportation capability}
In this section, we study whether a monogamy inequality like (\ref{eq:inequality}) exists for coherent state teleportation. In general, when entanglement measure $E$ is monotonically decreasing under discarding systems, there always exists a positive number $\alpha$ with which the inequality (\ref{eq:inequality}) is satisfied \cite{bib:AnnPhys.348.297}. The proof is simple. For given variables $x$, $y$, and $z$ satisfying $x>y>0$ and $x>z>0$, we can always find a positive number $\alpha$ giving
\begin{equation}
\left(\frac{y}{x}\right)^\alpha + \left(\frac{z}{x}\right)^\alpha < 1 .
\end{equation}
However, if $ \alpha $ tends to infinite, the inequality (\ref{eq:inequality}) becomes
\begin{equation} \label{eq:monotone}
\max \left\{ E_{A:B}, E_{A:C} \right\} \leq E_{A:BC} ,
\end{equation}
which is a trivial condition for entanglement monotone. In this case, it is not possible to find any faithful function $ f $ which describes the monogamy relation \cite{bib:PhysRevLett.117.060501}
\begin{equation}
f \left( E_{A:B}, E_{A:C} \right) \leq E_{A:BC} ,
\end{equation}
except for the trivial condition (\ref{eq:monotone}).

We employ a quantity called teleportation capability \cite{bib:PhysRevA.79.054309} defined as
\begin{equation}
\mathcal{C} = \max \{ 0, 2F^\textrm{opt}-1 \},
\end{equation}
which manifests quantum advantage beyond classical limit. We test whether teleportation capability $\mathcal{C}$ satisfies a monogamy inequality (\ref{eq:inequality}) and 
find that it does not hold for any finite order of $\alpha$. Below we explicitly construct an example which violates the monogamy inequality.

\subsection{Teleportation network}

First we need to define the teleportation fidelity $F_{A:B_1B_2\cdots B_N}^\textrm{opt}$ achievable when a sender has a single mode $A$ and a receiver has $N$ modes $B_1B_2\cdots B_N$ altogether. In this case, we may consider two different scenarios. One is teleportation network scheme where a teleportation is accomplished between two parties with the assistance of the others by local measurement and classical communication \cite{bib:PhysRevLett.84.3482}. The other scenario is to concentrate entanglement onto two-mode by global operations among $N$ modes \cite{bib:PhysRevA.71.032349} and to make use of concentrated two-mode entanglement for teleportation. It was shown that, for $(N+1)$-mode symmetric states, both scenarios yield the same optimal result \cite{bib:PhysRevLett.95.150503}. Here we briefly summarize the assisted teleportation scheme and its optimization.

We begin with one momentum-squeezed state and $N$ position-squeezed states, described by the quadrature operators in the Heisenberg picture
\begin{eqnarray}
& & \hat{x}_0 = e^{r_1}\hat{x}_0^{(n_1)}, \quad \hat{p}_0 = e^{-r_1}\hat{p}_0^{(n_1)} , \nonumber \\
& & \hat{x}_j = e^{-r_2}\hat{x}_j^{(n_2)}, \quad \hat{p}_j = e^{r_2}\hat{p}_j^{(n_2)} , (j = 1,2,\cdots,N) \nonumber \\
\end{eqnarray}
where the superscript $(n_a)$ refers to a thermal state with $ \left\langle (x^{(n_i)})^2 \right\rangle = \left\langle (p^{(n_i)})^2 \right\rangle = \frac{n_i}{2} $ for $ i = 1,2 $. Then we generate $(N+1)$-mode symmetric entangled states by combining the modes with beam splitter interactions
\begin{eqnarray}
& & \hat{B}_{N-1,N}\left(\cos^{-1}\frac{1}{\sqrt{2}}\right) \hat{B}_{N-2,N-1}\left(\cos^{-1}\frac{1}{\sqrt{3}}\right) \nonumber \\
& & \times\cdots\times \hat{B}_{0,1}\left(\cos^{-1}\frac{1}{\sqrt{N+1}}\right).
\end{eqnarray}
The mode $0$ belongs to Alice and the other $N$ modes to each of $N$ Bobs. Note that the entanglement is constant for fixed $n_1$, $n_2$, and $ \bar{r} = \frac{1}{2}(r_1+r_2) $. Therefore, two different states with the same $n_1$, $n_2$, and $ \bar{r} \equiv \frac{1}{2}(r_1+r_2) $ but with different $ d \equiv \frac{1}{2}(r_2-r_1) $ are equivalent under local operations and they can be transformed to each other by local squeezing operations. Similar to the standard two-mode protocol, Alice performs joint measurement of $\hat{x}_u$ and $\hat{p}_v$  and sends the outcomes to one of the receivers, namely $B_1$. On the other hand, $N-1$ other Bobs measure the momentum $p$ of their modes, respectively, and also send it to $B_1$. Then a displacement of $B_1$'s mode by
\begin{eqnarray}
\hat{x}_1 & \to & \hat{x}_\textrm{out} = \hat{x}_1+\sqrt{2}\hat{x}_u \nonumber \\
\hat{p}_1 & \to & \hat{p}_\textrm{out} = \hat{p}_1-\sqrt{2}\hat{p}_v + g\sum_{j=2}^N \hat{p}_j ,
\end{eqnarray}
accomplishes the teleporation, where $g$ is an adjustable gain.

The original proposal \cite{bib:PhysRevLett.84.3482} assumed a pure state with $ n_1 = n_2 = 1 $ and the same squeezing parameters $ r_1 = r_2 = \bar{r} $, with the argument that the scheme might be not optimal. The scheme was later optimized in \cite{bib:PhysRevLett.95.150503} with the squeezing parameters $r_1$ and $r_2$ adjusted for general cases with arbitrary $n_1$ and $n_2$. For optimal teleportation, fidelity is given by
\begin{eqnarray}
F_{A:B_1B_2 \cdots B_N}^\textrm{opt} & = & \frac{1}{1+2\nu_N} , \nonumber \\
\textrm{where } \nu_N & \equiv & \frac{1}{2}\sqrt{\frac{(N+1) n_1 n_2}{2 e^{4\bar{r}} + (N-1)\frac{n_1}{n_2}}} .
\end{eqnarray}
The quantity $\nu_N$ is exactly equivalent to the least symplectic eigenvalue of partially transposed CM under the bipartition $A:B_1B_2 \cdots B_N$ and the optimal fidelity thus has a direct relation to entanglement.
On the other hand, if we employ the standard two-mode teleportation between two modes $A$ and $B_i$, we find the optimal fidelity given by
\begin{eqnarray}
& & F_{A:B_i}^\textrm{opt} = \frac{1}{1+2\nu_1} , \nonumber \\
& & \textrm{where } \nu_1 \equiv \frac{1}{2}\sqrt{\frac{n_2}{N+1} \left( 2 n_1 e^{-4\bar{r}} + (N-1)n_2 \right)} .
\end{eqnarray}
The quantity $\nu_1$ here represents the least symplectic eigenvalue of partially transposed CM of two modes.

\subsection{Monogamy inequality for teleportation capability}

\begin{figure}[t]
\centering \includegraphics[clip=true, width=0.7\columnwidth]{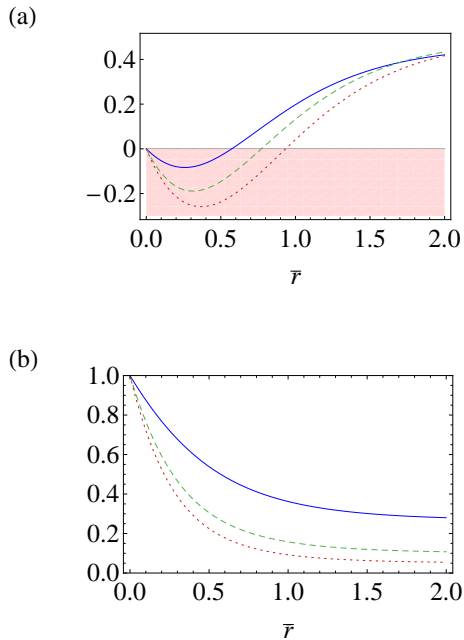}
\caption{\label{fig:inequality} Plot illustrating (a) $ \mathcal{C}_{A:B_1B_2 \cdots B_N}-\sum_{i=1}^N \mathcal{C}_{A:B_i} $ and (b) $ \frac{\mathcal{C}_{A:B_i}}{\mathcal{C}_{A:B_1B_2 \cdots B_N}} $ with respect to $\bar{r}$. Each curve corresponds to a different $N$: $N=2$ (blue solid), $N=5$ (green dashed), $N=10$ (red dotted). The shaded region represents the violation of monogamy inequality.}
\end{figure}

Let us consider a $N$-mode pure symmetric state with $n_1=n_2=1$ for a fixed $\bar{r}$. In this case, teleportation capabilities are given by
\begin{eqnarray}
\mathcal{C}_{A:B_1B_2 \cdots B_N} & = & \frac{2}{1+\sqrt{\frac{N+1}{2e^{4\bar{r}}+(N-1)}}} - 1 , \nonumber \\
\mathcal{C}_{A:B_i} & = & \frac{2}{1+\sqrt{\frac{2e^{-4\bar{r}}+(N-1)}{N+1}}} - 1 .
\end{eqnarray}
Note that $ \mathcal{C}_{A:B_i} > 0 $ regardless of the number of receivers $ N $, i.e., a sender can have quantum advantage beyond the classical bound with many different receivers. We plot $\mathcal{C}_{A:B_1B_2 \cdots B_N}-\sum_{i=1}^N \mathcal{C}_{A:B_i}$ for different values of $N$ in Fig. \ref{fig:inequality}(a), 
which shows a negative value for a small $\bar{r}$. In other words, teleportation capability does not satisfy the monogamy inequality with order $\alpha=1$.

We can also show that monogamy inequality is violated for any finite orders of $\alpha$. For small $\bar{r}<<1$, teleportation capabilities asymptotically behave as
\begin{eqnarray}
\mathcal{C}_{A:B_1B_2 \cdots B_N} & \approx & \frac{2\bar{r}}{N+1} + \frac{4(N-1)}{(N+1)^2}\bar{r}^2 + O(\bar{r}^3) , \nonumber \\
\mathcal{C}_{A:B_i} & \approx &\frac{2\bar{r}}{N+1} - \frac{4(N-1)}{(N+1)^2}\bar{r}^2 + O(\bar{r}^3) .
\end{eqnarray}
We thus see that
\begin{equation}
\mathcal{C}_{A:B_1B_2 \cdots B_N}^\alpha - \sum_{i=1}^N \mathcal{C}_{A:B_i}^\alpha \approx \frac{(N-1)2^\alpha}{(N+1)^\alpha} \bar{r}^\alpha \left( -1+2\alpha\bar{r} \right)
\end{equation}
becomes negative when $ \bar{r} \lesssim \frac{1}{2\alpha} $. In particular, in the small squeezing limit, the ratio of teleportation capabilities becomes $ \lim_{r \to 0} \frac{\mathcal{C}_{A:B_i}}{\mathcal{C}_{A:B_1B_2 \cdots B_N}} = 1 $ as shown in Fig. \ref{fig:inequality}(b). It means that every individual fidelity achieved between each pair is very close to the fidelity achievable collectively between Alice and all Bobs. In this case, a monotonic function $ f $ which satisfies $ f \left( \left\{ \mathcal{C}_{A:B_i} \right\} \right) \le \mathcal{C}_{A:B_1B_2 \cdots B_N} $ is only given by $ f \left( \left\{ \mathcal{C}_{A:B_i} \right\} \right) = \max \left\{ \mathcal{C}_{A:B_i} \right\} $, which is a trivial condition satisfied by entanglement monotone. Therefore, there does not exist any nontrivial monogamy inequality for CV teleportation capability in general.

\section{\label{sec:conclusion}Conclusion}

In summary, we investigated the CV teleportation using a multi-mode state to study monogamy property of useful entanglement. We showed that a strict monogamy relation holds for CV teleportation, i.e., a sender cannot beat no-cloning limit with more than one receiver, while it is possible to achieve quantum advantage beyond classical fidelity with any number of communicators. Starting with Gaussian entangled resources, one may attempt to improve teleportation fidelity individually using local Gaussian operations in collaboration with other parties, but it was shown to be not possible to beat the Gaussian no-cloning bound $ F_\textrm{nc}^G=\frac{2}{3}$. 
Even though the Gaussian no-cloning bound $ F_\textrm{nc}^G$ is slightly lower than the ultimate no-cloning bound, it is not even possible to overcome it readily by non-Gaussian operations acting on the Gaussian resource states. This provides a strong support to the strict monogamy relation in the CV QT.
On the other hand, we also showed that CV teleportation can achieve the optimal cloning, both Gaussian and non-Gaussian bound, by properly preparing the resource states, even if the QT generally constitutes a subset of all possible state manipulations considered in the no-cloning theorem. 

While one naturally expects a monogamy property of quantum entanglement qualitatively, it is a study of importance to identify a quantitative form of monogamy in order to look into the nature of quantum entanglement more deeply.
We further investigated the monogamy relation using an inequality form in terms of teleportation capability. We demonstrated that the monogamy inequality does not hold by constructing explicit examples. Nevertheless, a further study may be necessary to see if monogamy relation for QT can be described in a form of inequality accompanying an additional constraint. For example, it was shown that entanglement of formation or relative entropy of entanglement is not monogamous in general, but monogamy is recovered with dimension-dependent inequality \cite{bib:PhysRevLett.117.060501}. Since monogamy inequality for teleportation capability is violated in the regime of small $ r $, one may recover monogamy relation by taking additional constraint, e.g. energy constraint, into consideration.

\section*{acknowledgement}
We acknolwedge the support by an NPRP grant 8-352-1-074 from Qatar National Research Fund.

\end{document}